\documentclass[journal]{IEEEtran}

\ifCLASSINFOpdf
\else
  \usepackage[dvips]{graphicx}
  \graphicspath{{./}}
  \DeclareGraphicsExtensions{.eps}
\fi

\usepackage{url}

\begin{document}

%
\title{Performance of the ATLAS Liquid Argon Calorimeter after Three Years 
of LHC Operation and Plans for a Future Upgrade}
%
%
%

\author{Nikiforos~Nikiforou,
        on behalf of the ATLAS Collaboration
\thanks{N. Nikiforou is  with the Department of Physics, Columbia University, New York, NY 10027
USA (email: nikiforo@cern.ch). }
\thanks{Manuscript received on May 31, 2013.}}

%
%

\markboth{1304 INV}%
{Nikiforou \MakeLowercase{\textit{et al.}}: Performance of the ATLAS Liquid
Argon Calorimeter after three years of LHC operation and plans for a future
upgrade}
%


\maketitle
\thispagestyle{empty}

\begin{abstract}
The ATLAS experiment is designed to study the proton-proton collisions
produced at the Large Hadron Collider (LHC) at CERN. Liquid argon
sampling calorimeters are used for all electromagnetic calorimetry
as well as hadronic calorimetry in the endcaps. After installation in 2004--2006, the
calorimeters were extensively commissioned over the three--year period
prior to first collisions in 2009, using cosmic rays and single LHC
beams. Since then, approximately 27~fb$\mathbf{^{-1}}$ of data have been collected at an
unprecedented center of mass energy. During all 
these stages, the calorimeter and its electronics have been operating 
almost optimally, with a performance very close to specifications.
This paper covers all aspects of these first years of operation. 

The excellent performance achieved is especially presented in the context of the discovery of the
elusive Higgs boson. The future plans to preserve this performance until the end of the
LHC program are also presented.\end{abstract}


%
\IEEEpeerreviewmaketitle

\section{Introduction}
%
%
%
%
\IEEEPARstart{T}{he ATLAS} detector~\cite{Aad:2008zzm} is one of the two large
general-purpose experiments designed to study proton-proton as well as heavy-ion
collisions at the CERN Large Hadron Collider (LHC). Inside the LHC, bunches of up to $10^{11}$
protons collide nominally every 25~ns to provide proton-proton collisions at a
design luminosity of $\mathrm{10^{34}~cm^{-2} s^{-1}}$ and a center of mass energy up to 14~TeV.
For its first years of operation, the LHC has been operating at a reduced center of mass energy,
namely 7~TeV in 2011 and 8~TeV in 2012, and at an increased 50~ns bunch spacing. Already operating
close to its design luminosity, the LHC creates an extremely challenging environment for the
experiments, by producing multiple interactions per bunch crossing (\textit{pileup}), high radiation
doses and high particle multiplicities at unprecedented energies.

\begin{figure}[!t]
\centering
\includegraphics[width=3.5in]{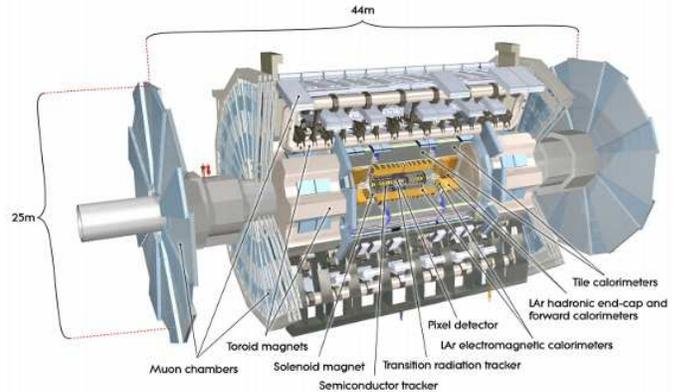}\caption{Cut-away view of the
ATLAS detector~\cite{Aad:2008zzm}.}
\label{fig:ATLASCutaway}
\end{figure}

The dimensions of the detector (Fig. \ref{fig:ATLASCutaway})  are 25~m in height and
44~m in length. The overall weight of the detector is approximately 7000~tonnes and it is housed in
an underground experimental cavern at Point--1 near the CERN main site. It covers nearly the entire
solid angle around the collision point, and successively consists of an Inner tracking Detector (ID)
surrounded
by a thin superconducting solenoid, electromagnetic (EM) and hadronic calorimeters, and a
 muon spectrometer incorporating three large toroidal magnet systems. 

\begin{figure}[!t]
\centering
\includegraphics[width=3.5in]{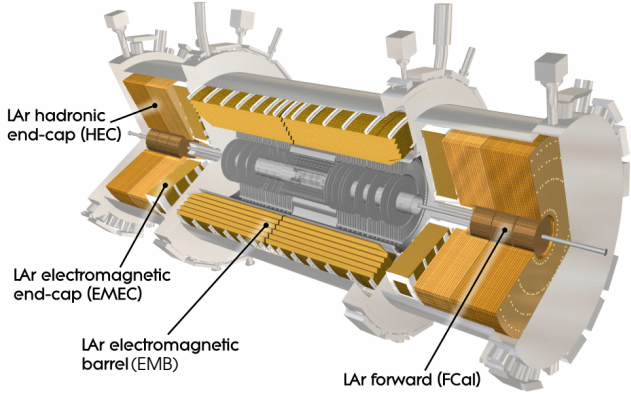}
\caption{The ATLAS LAr calorimeter system~\cite{Aad:2008zzm}.}
\label{fig:LAr_Labeled}
\end{figure}

The ATLAS calorimeter system covers the pseudorapidity\footnote{ATLAS uses a 
right-handed coordinate system with its origin at the nominal interaction point
(IP) in the center of the detector and the $z$-axis along the beam pipe. The
$x$-axis points from the IP to the center of the LHC ring, and the $y$-axis points
upward. Cylindrical coordinates $(r,\phi)$ are used in the transverse plane, $\phi$
being the azimuthal angle around the beam pipe. The pseudorapidity is defined in
terms of the polar angle $\theta$ as $\eta=-\ln\tan(\theta/2)$. In ATLAS, the so-called ``side-A''
refers to positive $z$-values while ``side-C'' refers to the negative side.} range $|\eta|< 4.9$ and
provides energy measurements of particles. Sampling calorimeters based on
Liquid Argon (LAr) technology are used for the detection of electromagnetic (EM) objects like
electrons and photons up to $|\eta| = 3.2$, as well as hadronic objects in the $|\eta|$ range
${1.5-4.9}$. The LAr calorimeter system (Fig.
\ref{fig:LAr_Labeled}) is the
subject of this paper and is described in more detail later. Hadronic calorimetry within $|\eta| <
1.7$ is provided by a steel/scintillator-tile calorimeter.

\section{The ATLAS LAr Calorimeter System}
In ATLAS, EM calorimetry is provided by barrel (${|\eta|<1.475}$) and endcap (${1.375<|\eta|<3.2}$)
accordion geometry lead/LAr sampling calorimeters. An additional thin LAr
presampler covering ${|\eta|<1.8}$ allows corrections for energy losses in material upstream of the
EM calorimeters.  The barrel electromagnetic (EMB) calorimeter~\cite{Aubert2006388}
consists of two half-barrels housed in the same cryostat.
The electromagnetic endcap calorimeter (EMEC)~\cite{AleksaEmec} comprises two wheels,
one on each side of the EM barrel. The wheels are contained in independent endcap cryostats
together with the hadronic endcap and forward calorimeters described later.
The wheels themselves consist of two co-axial wheels, with the outer wheel (OW)
covering the region ${1.375 < |\eta| < 2.5}$ and the inner wheel (IW) covering the region ${2.5 <
|\eta| < 3.2}$. 

The hadronic calorimetry provided by the tile calorimeter is complemented by two
parallel-plate copper/LAr hadronic endcap (HEC) calorimeters~\cite{1748-0221-2-05-P05005} 
that
cover the region ${1.5<|\eta|<3.2}$. Each HEC consists of two independent wheels which
combined provide 4 longitudinal calorimeter layers.

Finally, the forward calorimeters (FCal)~\cite{Artamonov:1094547} provide coverage over ${3.1
<|\eta|<4.9}$. In order to withstand the high particle fluxes in this region, they are based on a
novel design that uses cylindrical electrodes consisting of rods positioned concentrically inside
tubes parallel to the beam axis, supported by a metal matrix. Very narrow LAr gaps have been chosen
to avoid ion buildup at high rates and the gap is kept constant with a winding fiber wrapped around
the rods. Three cylindrical modules comprise the FCal, arranged sequentially; the module
closest to the IP is optimized for EM measurements and uses mainly copper as absorber and
$269~\mathrm{\mu m }$ gaps. The two subsequent modules are made mainly of tungsten and are optimized
for hadronic measurements with gaps of $375$ and $500 ~\mathrm{\mu m}$ respectively.

In total, the LAr calorimeter system comprises 8 subsystems (EMBA, EMBC, EMECA, EMECC, HECA, HECC,
FCalA and FCalC) collectively referred to as \textit{partitions}.

\subsection{Design and Principle of Operation}
\begin{figure}[!t]
\centering
\includegraphics[width=3.5in]{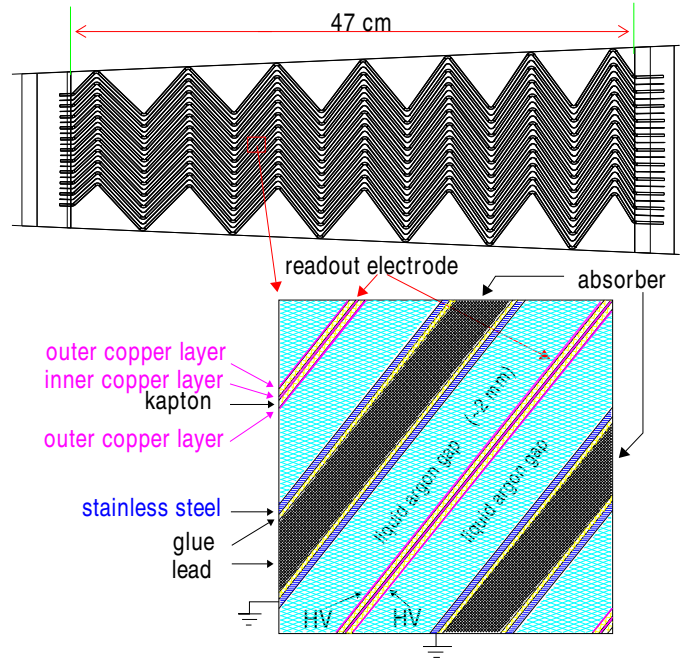}
\caption{Accordion structure of the barrel. The top figure is a view of a small sector of the
barrel calorimeter in a plane transverse to the LHC beams~\cite{driftimeMeasurement}.}
\label{fig:larg_principle}
\end{figure}

\begin{figure}[!t]
\centering
\includegraphics[width=3.5in]{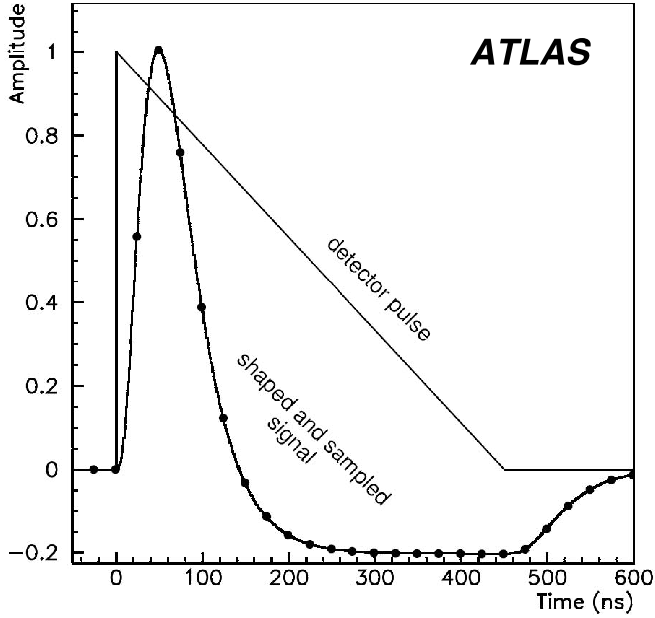}
\caption{Shapes of the LAr calorimeter current pulse in the detector and of the
signal output from the
shaper chip. The dots indicate an ideal position of samples separated by 25~ns~\cite{Aad:2008zzm}.}
\label{fig:shaperpulse}
\end{figure}

\begin{figure}[!t]
\centering
\includegraphics[width=3.5in]{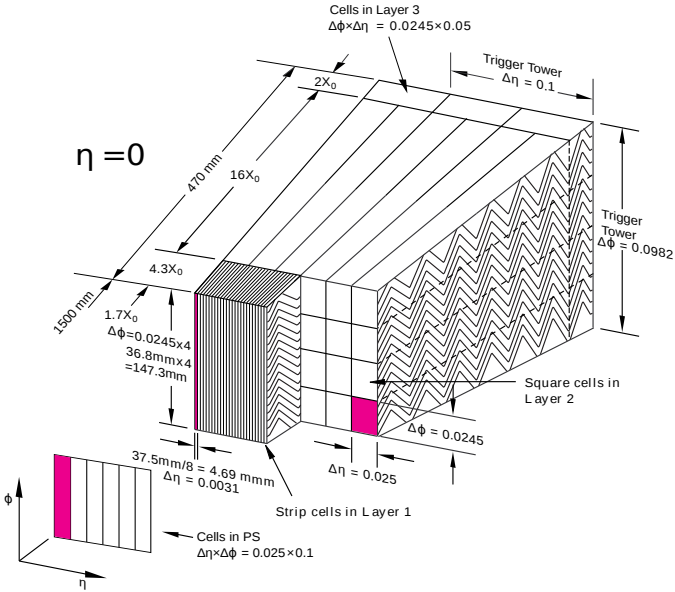}
\caption{Sketch of an EMB module where the different layers are visible. The
granularity in eta and phi of the cells of each of the three layers and of
the trigger towers is also shown~\cite{Aad:2008zzm}.}
\label{fig:LArSlice}
\end{figure}

The EM calorimeters comprise accordion-shaped copper-kapton electrodes positioned between lead
absorber
plates and kept in position by honeycomb spacers while the system is immersed in LAr (Fig.
\ref{fig:larg_principle}). Incident particles shower in the absorber material and subsequently the
LAr is ionized. Under the influence of the electric field between the grounded absorber and
powered electrode, the ions and electrons drift, the latter inducing a triangular pulse (Fig.
\ref{fig:shaperpulse}) to be collected by the electrodes. With the purpose of redundancy, both sides
of the electrodes are powered independently which allows for the collection of half of the signal
should one side lose power. In the EMB, the size of the drift gap on each side of the electrode
is 2.1 mm, which corresponds to a total electron drift time~\cite{driftimeMeasurement} of
approximately 450~ns for a nominal operating voltage of 2000~V. In the EMEC, the gap is a function
of radius and
therefore the HV varies with $\eta$ to provide a uniform detector response. To facilitate
installation, the absorbers and electrodes are ganged in $\phi$-modules. For most of the EM
calorimeter, EMB and EMEC-OW, each module has three layers in depth with different granularities, as
can be seen in Fig. \ref{fig:LArSlice}, while each EMEC-IW module has only two layers. The EM
calorimeter is designed so the largest fraction of the energy is collected in the second layer while
the back layer collects only the tail of the EM shower. Using the energy measurement and position
for all cells in all layers of the calorimeter contained in the shower, the incident particle energy
can be reconstructed and, taking advantage of the fine segmentation of the first layer, its
direction and characteristics can be inferred.  As discussed later, the fine segmentation is
extremely useful in the discrimination between photons and jets with a leading $\pi^0$ meson which
primarily decays to two photons. In addition, with its novel pointing geometry, the calorimeter can
reconstruct the direction of neutral particles, such as photons, for which semiconductor tracking
cannot be used.

The principle of operation is similar for the HEC
and FCal, though the design details, gap size, HV and drift time characteristics are different and
vary with position.

\subsection{Readout}
\begin{figure}[!t]
\centering
\includegraphics[width=3.5in]{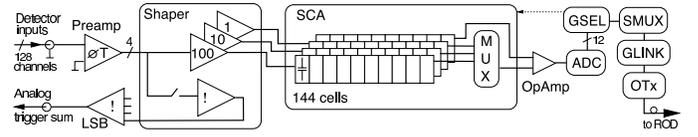}\caption{Front End Board block diagram
\cite{Aad:2008zzm}.}
\label{fig:ElecArchitecture}
\end{figure}

The ionization signals from all the cells are led outside the cryostats via 114 feedtroughs. Front
End Boards (FEBs)~\cite{1748-0221-3-09-P09003} housed in
crates mounted directly on the feedtroughs, receive the raw signals from up to 128 calorimeter
channels, process, digitize and transmit samples via optical link (see Fig.
\ref{fig:ElecArchitecture}) to the Back-End electronics housed outside the experimental cavern.  The
signal for each channel is split into three overlapping linear gain scales (Low, Medium and High) in
the approximate ratio 1/9/80, in order to meet the large dynamic range requirements for the expected
physics
signals. For each gain, the triangular pulse is shaped (Fig. \ref{fig:shaperpulse}) with a bipolar
$CR-(RC)^2$ analog filter to optimize the signal-to-noise ratio. The shaped signals are then
sampled at the LHC bunch-crossing frequency of 40 MHz and the samples for each gain are stored in a
Switched Capacitor Array (SCA) analog memory buffer while waiting for a {Level--1}  (L1) trigger
accept. For events accepted by the trigger, the optimal gain is selected for each channel, and the
samples are digitized and transmitted. In 2011 and 2012, typically 5 samples were digitized for each
pulse, whereas for the upcoming 14~TeV run, reducing the number of samples per pulse to 4 is being
considered.

In addition to the FEBs, the Front End Crates house several additional boards. Tower Builder Boards
facilitate the propagation of information to the trigger processor by summing calorimeter cells in
all layers in ``trigger towers'' with a size of approximately $0.1\times0.1$ in
$\Delta\eta\times\Delta\phi$. Calibration Boards allow the calibration of the electronics by
injecting a known exponential pulse to simulate the LAr ionization signal. The calibration signals
are then reconstructed through the regular readout chain.  Finally, auxiliary boards perform service
tasks such as clock distribution, communication and monitoring. 

\subsection{Cell Energy and Time Reconstruction}
The optimal filtering~\cite{Cleland:1072138} technique is used to reconstruct the cell energy and
peaking time from the samples of the shaped calorimeter pulse. The procedure described here applies
to all LAr subsystems though it differs slightly in the case of the FCal.  To calculate the cell
energy, $E_{\mathrm{cell}}$ in MeV, from the samples $s_j$ in ADC counts, the following formula is
used: 
\begin{displaymath}
E_{\mathrm{cell}} = F_{\mu A\rightarrow MeV}\cdot F_{DAC\rightarrow\mu A}\cdot
\frac{1}{\frac{M_{\mathrm{phys}}}{M_{\mathrm{cali}}}}\cdot R \sum_{j=1}^{N_{\mathrm{samples}}}
a_j
\left(s_j - p\right)
\end{displaymath}
while to calculate the time a similar formula is used:
\begin{displaymath}
t_{\mathrm{cell}} = \frac{1}{E_{\mathrm{cell}}}\sum_{j=1}^{N_{\mathrm{samples}}} b_j
\left(s_j - p\right)
\end{displaymath}
where $F_{\mu A\rightarrow MeV}$ is a coefficient that is obtained from test
beam studies and converts the ionization current values to energy values, $F_{DAC\rightarrow\mu A}$
is a property of the calibration board and $\frac{M_{\mathrm{phys}}}{M_{\mathrm{cali}}}$ is a
factor to correct for differences between the physics signal and calibration pulses. $R$ is the
ramp slope and $p$ is the pedestal (electronic baseline) obtained from calibration. The parameters
$a_{j}$ and $b_j$ are sets of  Optimal Filtering Coefficients calculated from the knowledge of
the calibration pulse shape and the noise autocorrelation function, to give the optimal energy and
time resolution. Finally, a \textit{Quality Factor}, $Q^2$, is calculated for each cell, as an
estimate of the quality of the reconstructed pulse.

\section{Operation}
In order to meet the performance requirements, a significant effort is made by the LAr
Operations group to continuously monitor the detector status and performance and take corrective
actions if needed. This task is supported by various monitoring systems conceived and installed for
this purpose. A Detector Control System (DCS) has been developed in the ATLAS-wide DCS framework
to provide control, monitoring and human interface with the hardware, based on the commercial
SCADA software PVSS-II (now named SIMATIC WinCC Open Architecture)~\cite{etmWebsite}.

\subsection{Cryogenic System Stability}
The cryogenic system aims to provide stable LAr conditions. The
temperature sensitivity of the LAr calorimeter has been determined to follow a linear relationship
with a 2\% decrease of the measured signal per kelvin. This includes the contribution from LAr
density variations with temperature as well as drift velocity variations, contributing to
changes of -0.45\%/K and -1.55\%/K respectively. In order to meet the design energy resolution,
 a 100 mK temperature stability and uniformity is required.

A temperature monitoring system is in place to monitor and log the conditions so as to be able to
correct for any temperature variations and non-uniformities. The temperature monitoring system
comprises more than 400 calibrated precision temperature probes (PT100 platinum resistors)
distributed throughout the three cryostats and immersed in the LAr. Measurements are taken every
minute and a conditions database is updated if there is a change from the previous measurement.  

\begin{figure}[!t]
\centering
\includegraphics[width=3.5in]{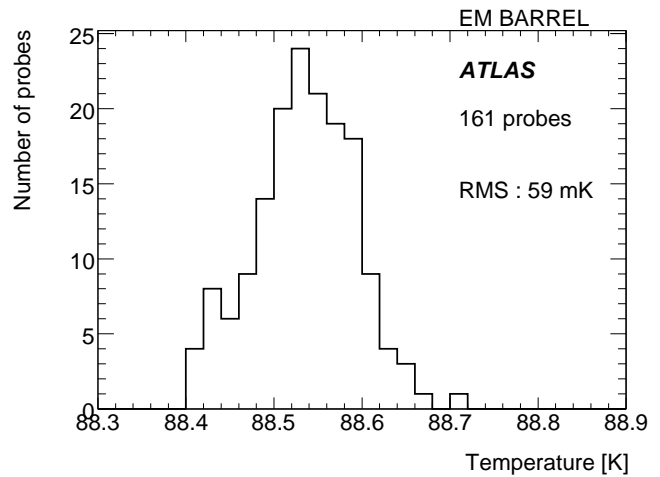}
\caption{Distribution of the average temperature measured over ten days for temperature probes
within the EM Barrel cryostat~\cite{stab}.}
\label{fig:barrelTemp}
\end{figure}

\begin{figure}[!t]
\centering
\includegraphics[width=3.5in]{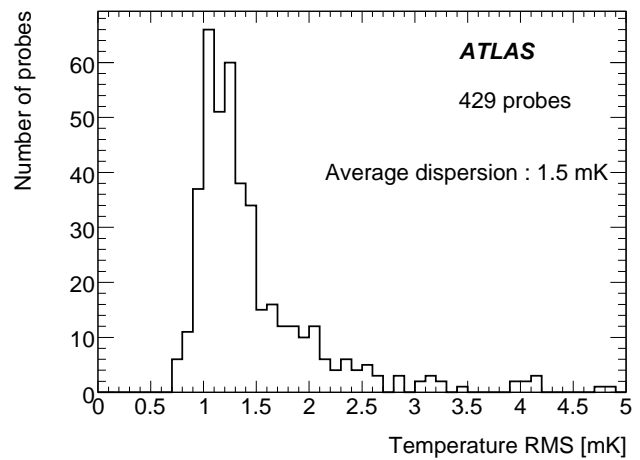}
\caption{Distribution of the RMS of the temperature measurements, over ten days, for each of the
temperature probes within all three of the LAr calorimeter cryostats~\cite{stab}.}
\label{fig:barrelRMS}
\end{figure}

Periodic studies are performed to monitor the stability of the cryogenic system during data taking.
Temperature data are collected over long periods of time without interventions on
the cryogenic system and low voltage power supply and for constant general conditions, such as the
magnetic field. Fig. \ref{fig:barrelTemp} shows the temperature uniformity of the LAr barrel
calorimeter, demonstrating an overall barrel temperature uniformity of 59 mK for data collected
over a period of 10 days. Similar or better results are obtained in the endcaps, satisfying the
design requirements. Finally, Fig. \ref{fig:barrelRMS} shows the distribution of the RMS of the
temperature measurements for each of the temperature probes in the LAr calorimeter cryostats, over
the same period. The average RMS of 1.5~mK demonstrates the reliability of the temperature
control and monitoring system.

An additional source of non-uniformity in the cryogenic system arises from the presence of
impurities in the LAr. Ionization signal electrons can attach to electronegative
impurities, such as $O_2$, which would lead to the degradation of the signal measurement. A
limit of 1000~ppb oxygen impurities or equivalent has been set for the efficient operation of the
calorimeter.

\begin{figure}[!t]
\centering
\includegraphics[width=3.5in]{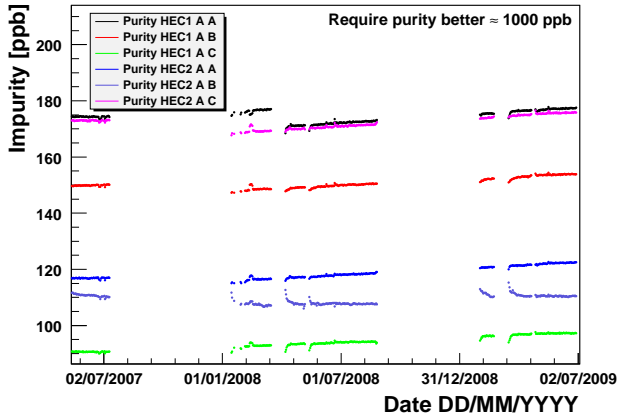}
\caption{Impurity levels in $O_2$ equivalents for HECA. Each point shows the
mean of the purity values measured over a single day and excludes extreme or unreasonable values
that can be attributed to instrumental effects~\cite{stab}.}
\label{fig:purity}
\end{figure}

The stability of the LAr purity is monitored in 10 to 15 minute intervals by a system of 30
purity monitors~\cite{Adams:2005xh} installed throughout the cryostats. Each purity monitor measures
the deposition of known charges from two contained monoenergetic radioactive sources ($^{207}Bi$
and $^{241}Am$) in ionization chambers. The monitors were shown to have a systematic uncertainty of
19\% and a statistical uncertainty for a single measurement of $\sim 5$~ppb.  
Fig. \ref{fig:purity} shows the impurity levels over the course of two years between July 2007
and July 2009 for HECA. The overall levels of impurities have been measured to be approximately
$200$~ppb and $140$~ppb in the barrel and endcap cryostats respectively, well below the operational
limit. Finally, the levels have been shown to be stable except in the case of side C of the endcap
where a small degradation of approximately $5$~ppb per year is observed.

\subsection{High Voltage Operation}
The high voltage (HV) system provides the electric field so that the ionization signal
propagates to the measurement electrodes for readout. The HV values are monitored in real time
via the DCS system and stored in a conditions database. Suitable corrections are applied during the
energy reconstruction for any deviations from the nominal values. 

\begin{figure}[!t]
\centering
\includegraphics[width=3.5in]{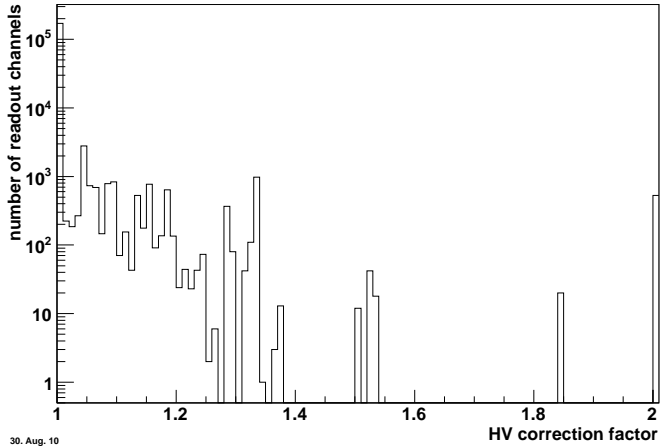}
\caption{Distribution of cell energy correction factors applied for channels serviced by at least
one HV line operated at reduced voltage~\cite{pubPlots}.}
\label{fig:hvcorrfactors}
\end{figure}

During data taking, some HV hardware modules may trip, which leads to loss of power to one side of
an electrode, thereby inhibiting the signal measurement. Procedures have been established for the
manual recovery of tripped HV lines on a case-by-case basis. In addition, an autorecovery system has
been implemented for modules that are not tripping systematically. Owing to the frequent measurement
of the HV values, data recorded during the ramp-up of tripped lines can be corrected and used in
analysis. Specialized studies were performed to verify that data collected during HV
recovery can safely be used in analysis. During
the trip itself, the rapid variation of the voltage makes the proper correction of the signal
impractical, leading to a small loss of data. In order to limit the number of trips and minimize
this
inefficiency, problematic lines that trip frequently are set to lower operating HV values and a
correction is applied for the energy of the affected cells. A typical distribution of correction
factors is shown in Fig. \ref{fig:hvcorrfactors}. Finally, more robust HV modules have been deployed
that allow the module to run in current control mode instead of tripping, which prevents the voltage
drop. All the above improvements result in the minimization of data rejection due to HV trips in
2012 which was limited to 0.46\%.

\subsection{Electronics Calibration and Stability}\label{sec:calib}
Calibration campaigns are performed regularly and involve special runs in the absence of LHC beams
to measure the response of
the FEB electronics in situ for each of the three gains. An
infrastructure has been deployed to facilitate the retrieval and processing of those runs, the
extraction of calibration constants and their propagation to a conditions database. 

Daily pedestal and noise levels for each channel are determined by ``Pedestal'' runs where the
channel is triggered without signal injection. The pedestal is computed for each channel from the
average over the triggers and the number of samples.

In order to determine the gain for each channel, ``Ramp'' runs are also taken on a daily
basis. In these runs, each channel is pulsed several times with a set of predetermined calibration
board DAC values, corresponding to input currents. For each DAC value, the resulting pulses
are reconstructed and an average pulse shape is computed from which the peak ADC value is
measured. Thus, the relationship between the DAC and ADC values is obtained for each
channel and, by fitting it with a first order polynomial, the ramp slope $R$ is calculated. 

\begin{figure}[!t]
\centering
\includegraphics[width=3.5in]{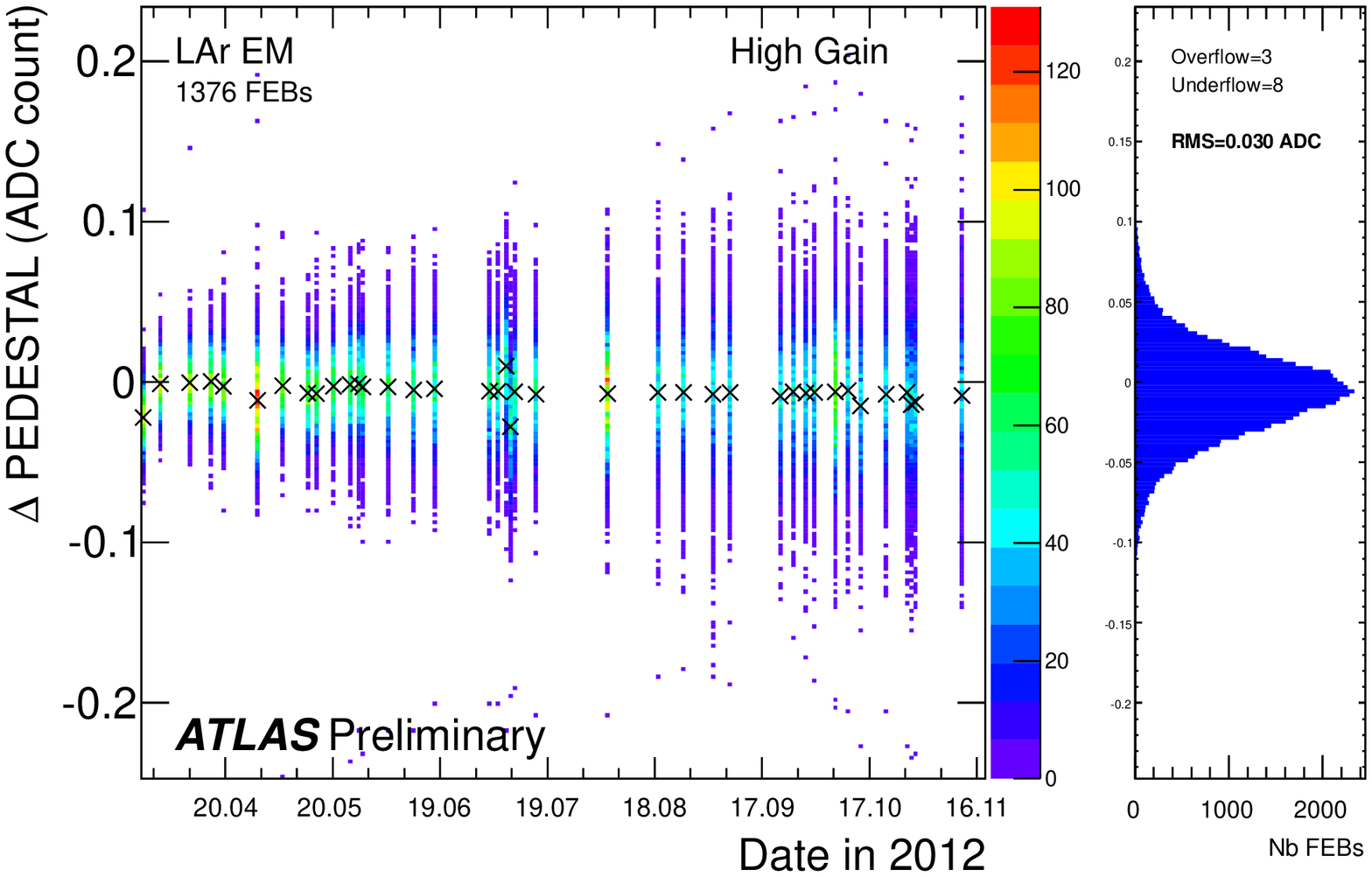}
\hfil
\includegraphics[width=3.5in]{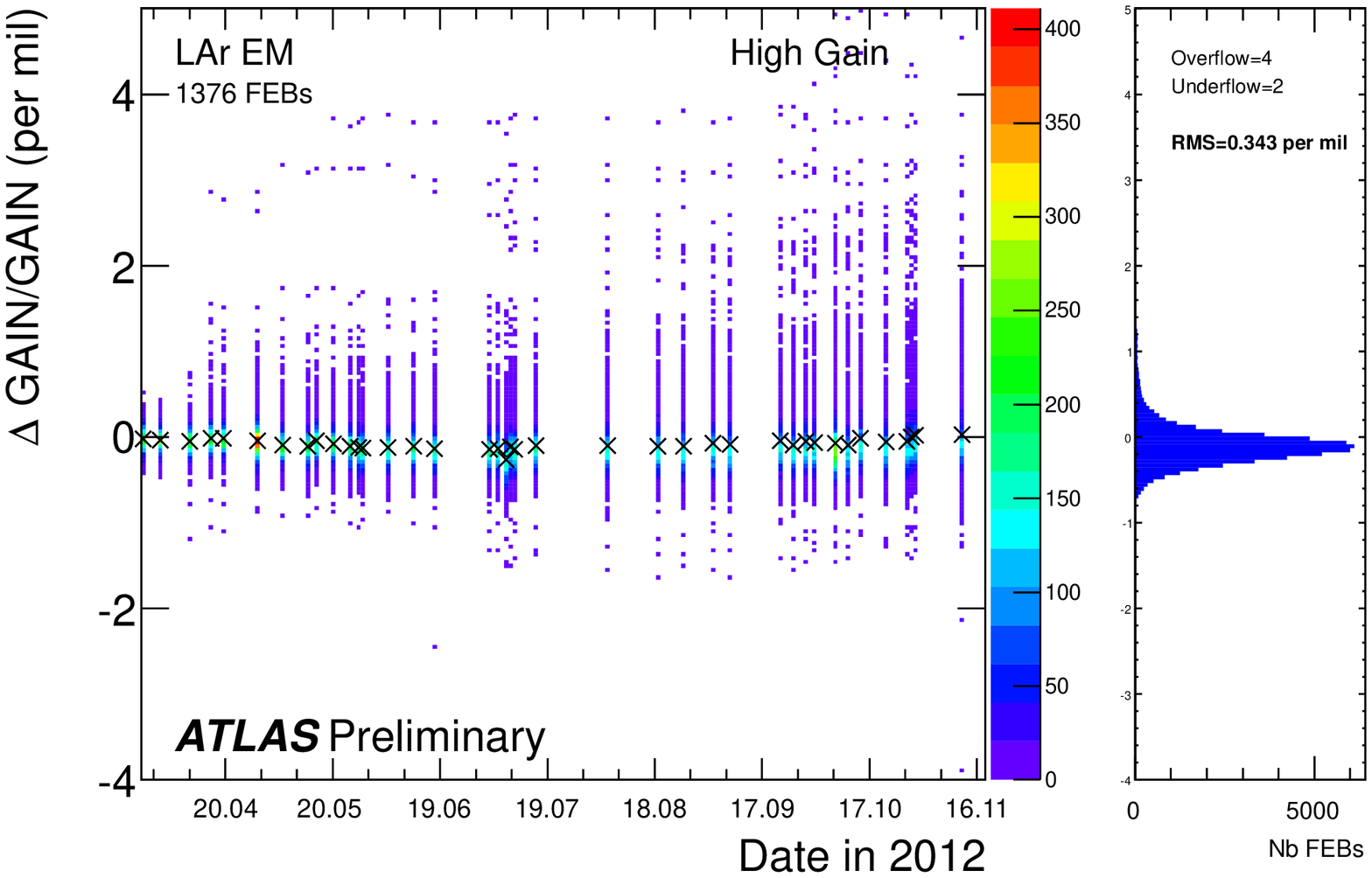}
\caption{Pedestal stability (top) and ramp stability
(bottom) with respect to time in 2012 for the LAr EM calorimeters in High gain~\cite{stab}.}
\label{fig:calibStability}
\end{figure}

To demonstrate the excellent stability of the LAr electronics, Fig. \ref{fig:calibStability} shows
example stability plots for the pedestal and gain values with respect to time in 2012, for FEBs in
the LAr EM calorimeters (EMB and EMEC) in High gain. For each FEB, the pedestal and gain average
over all channels serviced by that board is calculated and plotted as a function of time. Compared
to a reference value, the pedestal difference distribution has an RMS of 0.030 ADC counts, while the
relative gain difference distribution has an RMS of 0.343 per mil. The stability is similar for the
other LAr calorimeters (HEC and FCal) and in the other gains, with a pedestal difference RMS ranging
from 0.022 to 0.029 ADC counts. Finally, the relative gain difference is distributed with an RMS
between 0.316 and 0.343 per mil for the EM calorimeters, 0.657--0.814 per mil for the HEC and
0.046--0.086 per mil for the FCal. 

Finally, ``Delay'' runs are taken weekly or as often as it is needed to reconstruct the pulse shape
itself in detail. In this mode, for each calibration pulse, the channel is triggered several times 
with incremental delays of 1.04~ns. On each trigger, 32 samples are acquired at 40~MHz and the
entire pulse is therefore effectively sampled at 1.04~ns, allowing reconstruction of the pulse shape
with high precision.  Each channel  is further pulsed several times and the procedure is repeated
to calculate an average pulse shape. The shape is used to extract the Optimal Filtering Coefficients
that describe the  shape and are used in the energy and time reconstruction from pulse samples as
well as the calculation of the quality of the reconstructed pulse.              

The values of the calibration constants are monitored for significant variations and are updated in
the database typically once a month, or whenever there is a change in the system conditions.

\subsection{Timing}
A precise timing measurement of calorimeter signals is particularly useful in the operation of
the detector as well as physics analyses. Knowledge of time with a precision small compared to 25~ns
(the nominal LHC bunch interval) is required to differentiate between energy deposits originating
from collisions in the triggered bunch crossing as opposed to signals from neighboring bunch
crossings. The latter phenomenon is known as ``out-of-time pileup''. Further,
excellent timing can be used to reject other sources of beam induced background such as satellite
collisions and beam halo~\cite{ATLAS-CONF-2011-137}. In addition, it provides a means to veto events
with energy deposited by cosmic ray interactions
overlapping with the triggered event. Finally, the timing can be employed in searches for new
physics involving long-lived neutral particles decaying to photons~\cite{Aad:2013oua}, electrons or
jets.

\begin{figure}[!t]
\centering
\includegraphics[width=3.6in]{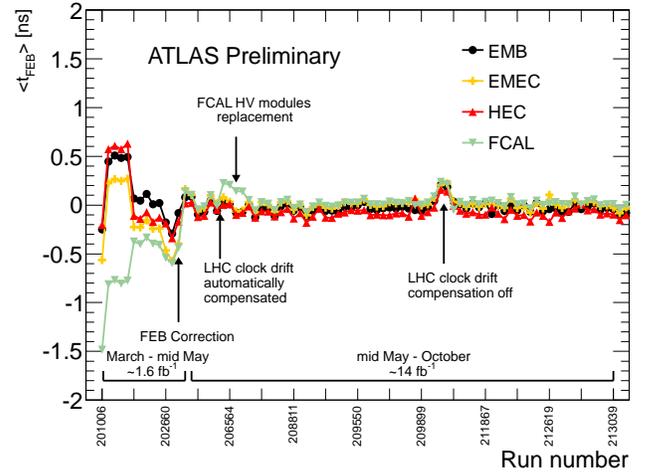}
\caption{Mean FEB time synchronization for each LAr partition~\cite{pubPlots}.}
\label{fig:FEBTimeEvolution}
\end{figure}

The timing measurement is performed with respect to the LHC clock phase delivered from a central
system to ATLAS by means of an optical fiber. The clock phase can drift, for example, due to
variations of the fiber length with temperature, and this can be detected by comparing with the
actual bunch arrival times in ATLAS. The average time is monitored and is corrected for any global
variations such as the LHC clock drift. In 2012, the capability to automatically compensate
for any clock drift with a precision of a few tens of ps was implemented during the June technical
stop and performed very well as shown in Fig.~\ref{fig:FEBTimeEvolution}.

\begin{figure}[!t]
\centering
\includegraphics[width=3.6in]{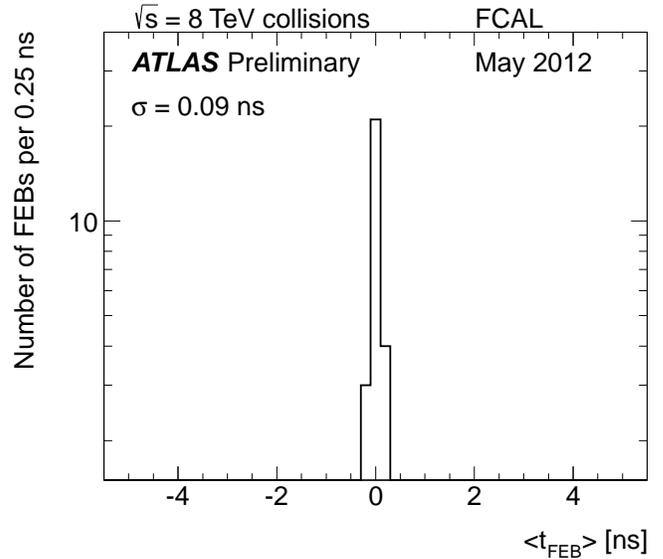}
\caption{Average time per Front End Board in the FCal with 8~TeV collision data on May 2012
\cite{pubPlots}.}
\label{fig:t_FEB_FCAL}
\end{figure}

The calorimeter is timed in with collision data with an online precision of approximately 1~ns. As
shown in Fig. \ref{fig:t_FEB_FCAL}, the calorimeter is properly aligned and uniform in time, with a
typical spread of the average FEB time between 0.09~ns and 0.17~ns, depending on the partition.

\begin{figure}[!t]
\centering
\includegraphics[width=3.5in]{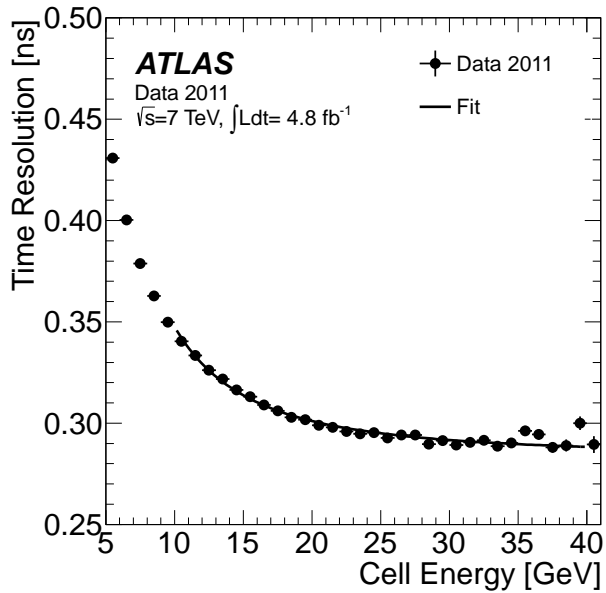}
\caption{Single cell time resolution energy dependence, for cells reconstructed in High gain, in
the Middle layer of the LAr EM Barrel Calorimeter (${|\eta|<1.4}$). A fit with the expected energy
dependence, comprising a so-called ``noise term'' and a ``constant term'', added in quadrature, is
shown superimposed~\cite{Aad:2013oua}.}
\label{fig:timingres}
\end{figure}

To improve the timing performance of the EM calorimeters further, a number of effects were studied
and corrected for with good quality ${W\rightarrow e\nu}$ events. Additional calibration constants
are computed as a function of the data taking period, FEB, channel within the FEB, cell energy
and primary vertex position. After all constants are applied, a timing
resolution of $\approx 290$~ps is achieved for large energy depositions in the EMB, as shown in Fig.
\ref{fig:timingres}. By comparing the corrected time of the two electrons in ${Z\rightarrow ee}$
events, this resolution is understood to include a correlated contribution of $\approx220$~ps,
expected to be dominated by the spread of the proton bunches along the LHC beamline, and an
uncorrelated contribution of $\approx 190$~ps. The latter component includes the intrinsic timing
resolution of the LAr EMB and its readout, as well as residual non-uniformities and imperfections in
the calibration procedure.

\section{Data Quality Monitoring}\label{sec:dqm}
A complex monitoring procedure
was designed to track and quickly identify all potential problems
affecting the detector, including the situations described in previous sections. The Data Quality
Monitoring effort involves low level hardware monitoring via the DCS system as well as checks
in every stage of the data acquisition and processing chain. Several algorithms assist trained
personnel during the data taking (online monitoring) in detecting issues that could compromise data
quality, so corrective actions can be taken. Offline, more detailed tests are performed using a
subset of the recorded data which are expeditiously reconstructed to pinpoint blocks of data with
suspected quality issues, or \textit{defects}, before the processing of the bulk of the data is
launched 48 hours later. The results of the tests are organized through a dedicated web
infrastructure and trained experts decide on the defect type, localization, severity and length in
time, with the ultimate goal of optimal detector efficiency. During this process, the detector
conditions during the data taking, stored via DCS are also taken into account (for example HV
values and status). The proposed actions are propagated as needed to the relevant conditions
databases via automated procedures. After the bulk processing is completed, typically several days
later, a final data quality assessment is performed to verify that the problems identified in the
previous steps have been handled properly. If needed, the databases are further updated to flag
the remaining issues and, if possible, to allow the recovery of data in future reprocessing. 

\begin{figure}[!t]
\centering
\includegraphics[width=3.6in]{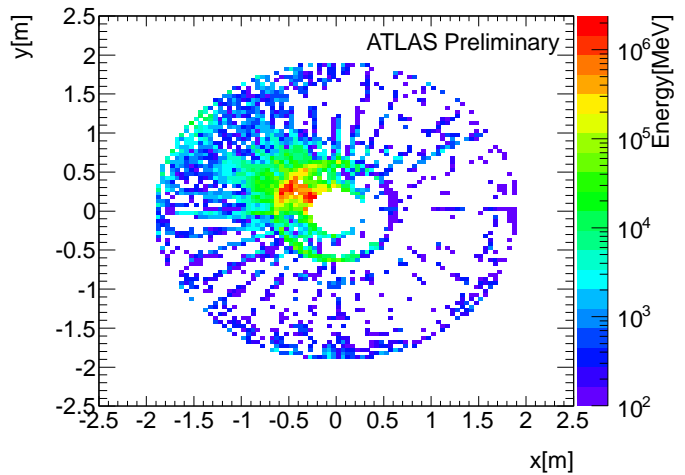}
\caption{Example of a single event affected by a noise burst recorded in empty LHC bunches
in EMECA~\cite{pubPlots}.}
\label{fig:noiseBurstEvent}
\end{figure}

A significant source of possible LAr inefficiency with an interesting pathology is the presence of
bursts of large scale coherent noise, or \textit{noise bursts}, mainly located in the endcaps. This
phenomenon manifests itself only in the presence of collisions and was found to scale with
instantaneous luminosity. The effect is very short in time, lasting usually less than
$\sim5~\mathrm{\mu s}$, and during that time a significant percentage of channels exhibit signal
readings which are significantly above the typical electronic noise levels.  An example of such an
event is shown in Fig. \ref{fig:noiseBurstEvent}.  

\begin{figure}[!t]
\centering
\includegraphics[width=3.6in]{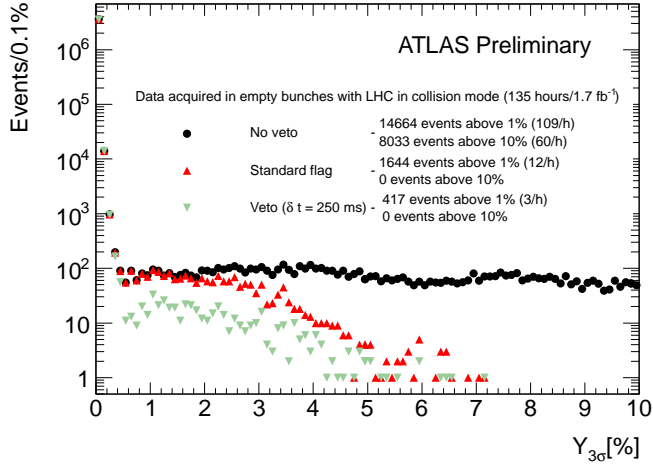}
\caption{Distribution of the percentage of noisy channels in the EM
endcap, measured during empty bunches in LHC collision mode. The black points show the percentage
without any corrective action. The red points represent the data after rejection of events
identified by means of the quality factor only while the green points show the effect of the time
veto procedure~\cite{pubPlots}.}
\label{fig:noiseBurstCleaning}
\end{figure}

A useful variable for the description of noise burst events is $Y_{3\sigma}$ which represents the
percentage of channels with a signal greater than three times the electronic noise measured during
the LHC beam crossings without collisions (empty bunches). Hard noise burst events with large
$Y_{3\sigma}$ are efficiently identified and flagged using the quality factors of the pulse
measurements. However it was found that softer noise bursts, usually peripheral to a hard event and 
characterized by a $Y_{3\sigma}$ of the order 2-3\%, are not efficiently flagged. Taking advantage
of the short nature of the phenomenon, a \textit{time window veto} procedure was established to
identify these occurrences and reject neighboring events within a conservative time window around
the identified noise burst. The window length was 1~s
for 2011 and 250~ms for the 2012 data taking period. The efficiency of this method in rejecting
large scale coherent noise events in empty bunches is demonstrated by Fig.
\ref{fig:noiseBurstCleaning}.

The procedures described above were established and improved over the last few years of ATLAS
operation and resulted in a continuous improvement in data quality and near optimal LAr efficiency.
For 2012, the LAr inefficiency was limited to 0.88\% of which 0.46\% is due to HV trips while
0.2\% is attributed to noise bursts.  

\section{Physics Performance}
While the low-level detector hardware and Data Quality monitoring are paramount to the successful
running of the experiment, the ultimate goal is to reconstruct interesting physics events with
efficiency, reliability and precision. It is the responsibility of the
LAr calorimeter system to provide excellent measurements of physics objects, such as electrons,
photons and jets, and contribute to the calculation of the event missing energy with high
resolution. Tests at the physics level are performed continuously during the data taking periods
using benchmark physics processes to gauge the object reconstruction stability and precision. 

\subsection{Energy Scale, Resolution and Stability}
A detailed study was performed~\cite{egamma} to determine the electron performance of the ATLAS
detector using the decays of the $Z$, $W$ and $J/\psi$ particles. The studies demonstrate the
level of energy scale, resolution and uniformity of the LAr system as well the excellent overall
performance of the ATLAS detector.

\begin{figure}[!t]
\centering
\includegraphics[width=3.6in]{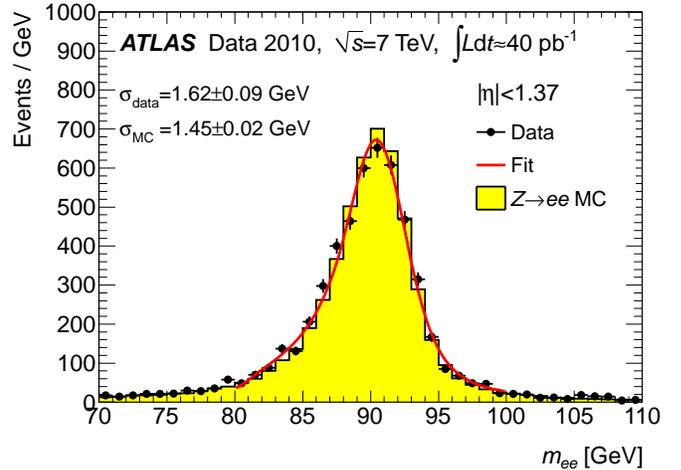}
\caption{Reconstructed dielectron mass distribution for $Z\rightarrow ee$ decays for $|\eta|< 1.37$
and comparison to simulation~\cite{egamma}.}
\label{fig:zeeMassBarrel}
\end{figure}

The EM showers that develop in the calorimeters are reconstructed as clusters of calorimeter cells
that contain a large fraction of the deposited energy. Some energy is not contained in the
cluster and some is lost before or after the calorimeter. For these reasons, corrections need to be
applied. Calibration constants are calculated from MC simulation as a function of
$\eta$, energy and shower depth. The overall energy scale is set with reconstructed mass
distributions from $Z\rightarrow ee$ events (see Fig. \ref{fig:zeeMassBarrel} for an example in the
barrel) and cross-checked using the electron $E/p$ distribution in $W\rightarrow e\nu$ events. The
latter takes advantage of the independent measurements of the electron energy, $E$ in the
  calorimeter and its momentum, $p$ in the Inner Detector (ID). 

\begin{figure}[!t]
\centering
\includegraphics[width=3.6in]{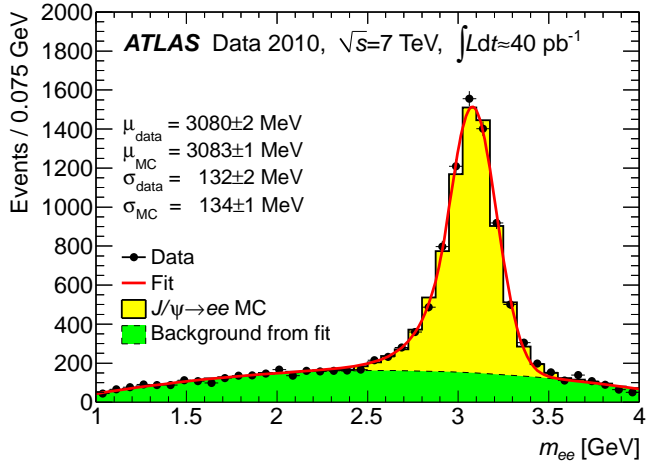}
\caption{Reconstructed dielectron mass distribution for $J/\psi\rightarrow ee$ decays and
comparison to simulation~\cite{egamma}.}
\label{fig:jpsiMass}
\end{figure}

Having set the energy scale using the calibration constants, the calorimeter energy resolution can
be studied. The resolution is usually parametrized with the following formula:
\begin{displaymath}
 \frac{\sigma_E}{E} = \frac{a}{\sqrt{E}} \oplus \frac{b}{E} \oplus c
\end{displaymath}
where $a$, $b$ and $c$ are $\eta$-dependent parameters called the \textit{sampling term},
\textit{noise term} and \textit{constant term} respectively. Mass measurements in dielectron
events, are compared to simulation and the lineshape of the $J/\psi$ decay is used to verify that
the
sampling and noise terms are modeled correctly in simulation (Fig. \ref{fig:jpsiMass}). The width of
the $Z\rightarrow ee$ decay mass distribution is then used to extract an
\textit{effective constant term}, $c_{\mathrm{{data}}}$, including the calorimeter constant
term as
well as the effect of inhomogeneities and residual mis-calibration. For each $\eta$-range,
the following formula is used:
\begin{displaymath}
 c_{\mathrm{data}} =
\sqrt{2\cdot\left(\left(\frac{\sigma}{m_{Z}}\right)^2_{\mathrm{data}}-\left(\frac{\sigma}{m_{Z}}
\right)^2_{\mathrm{MC}}\right)+c^2_{\mathrm{MC}}}
\end{displaymath}
where $c_{\mathrm{MC}}$ is the constant term in simulation, while $m_Z$ denotes the $Z$ mass and
$\sigma$ the Gaussian component of the experimental resolution, obtained by fitting with a
Breit-Wigner convolved with a
Crystal Ball function~\cite{CrystalBall1,CrystalBall2}. Recent results for all $\eta$-ranges are
shown in Table \ref{resTable}.

\begin{table}[!t]
\renewcommand{\arraystretch}{1.3}
\caption{Calorimeter Energy Resolution Constant Term for all partitions}
\label{resTable}
\centering
\begin{tabular}{c||c||c}
\hline
\bfseries Subsystem & \bfseries $\eta$-range & \bfseries Effective constant term,
$c_{\mathrm{data}}$\\
\hline\hline
EMB & $|\eta|<1.37$ & $1.2\%\pm 0.1\%\mathrm{(stat)}{}^{+ 0.5\%}_{-0.6\%}\mathrm{(syst)}$\\
EMEC-OW & $1.52<|\eta|<2.47$ & $1.8\%\pm 0.4\%\mathrm{(stat)}\pm 0.4\%\mathrm{(syst)}$\\
EMEC-IW & $2.5<|\eta|<3.2$ & $3.3\%\pm 0.2\%\mathrm{(stat)}\pm 1.1\%\mathrm{(syst)}$\\
FCal & $3.2<|\eta|<4.9$ & $2.5\%\pm 0.4\%\mathrm{(stat)}{}^{+ 1.0\%}_{-1.5\%}\mathrm{(syst)}$\\
\hline
\end{tabular}
\end{table}

\begin{figure}[!t]
\centering
\includegraphics[width=3.5in]{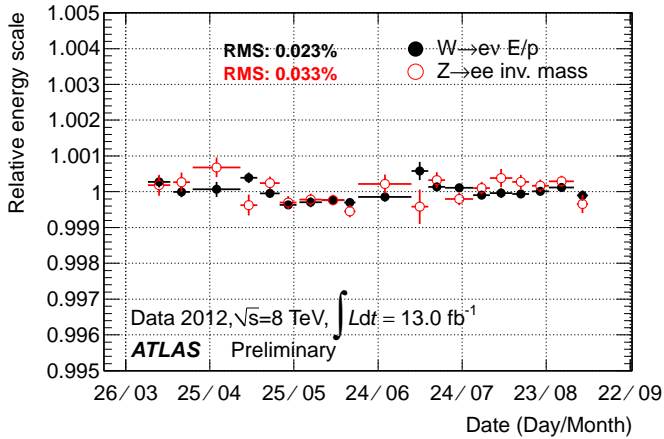}
\caption{Electron energy response stability with time in 2012 data~\cite{Teischinger:1491209}.}
\label{fig:TimeStability}
\end{figure}

\begin{figure}[!t]
\centering
\includegraphics[width=3.6in]{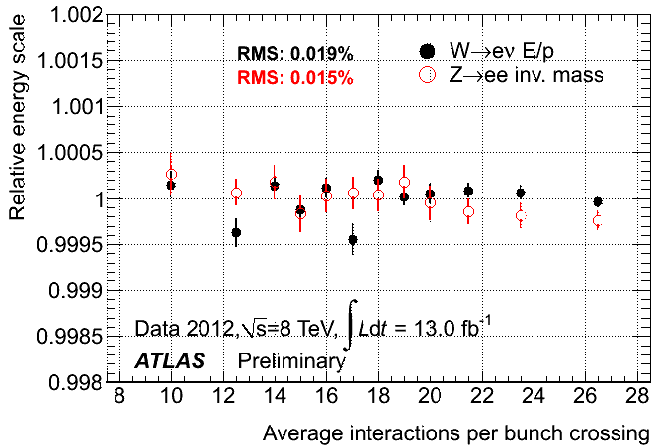}
\caption{Electron energy response stability with pileup (average number of interactions per bunch
crossing) in 2012 data~\cite{Teischinger:1493595}.}
\label{fig:MuStability}
\end{figure}

Finally, the electron energy response stability is demonstrated with respect to time
(Fig. \ref{fig:TimeStability}) as well as the LHC pileup conditions (Fig. \ref{fig:MuStability})
by studying the evolution of the peak of the invariant mass distribution in $Z\rightarrow ee$
events and the most probable value of the $E/p$ ratio in $W\rightarrow e\nu$ events, with respect
to a reference value. The stability in 2012 is excellent, better than $0.03\%$ which is
expected for a LAr calorimeter with stable temperature and purity conditions. 

\subsection{Identification Efficiency}\label{sec:id}
In addition to having good resolution and uniformity, the calorimeter is required to have excellent
electron/photon identification efficiency and a high jet rejection rate over a broad energy range.
The fine segmentation of the LAr calorimeter makes this possible by providing valuable information
for the shape and other characteristics of the EM showers. In addition, photons which leave no
tracks in the ID can be efficiently detected and their direction measured using the same feature.
More specifically, the photon flight direction can be reconstructed by measuring precisely the
position of the cluster barycenters in the first and second LAr calorimeter layers in depth.

\begin{figure}[!t]
\centering
\includegraphics[width=3.6in]{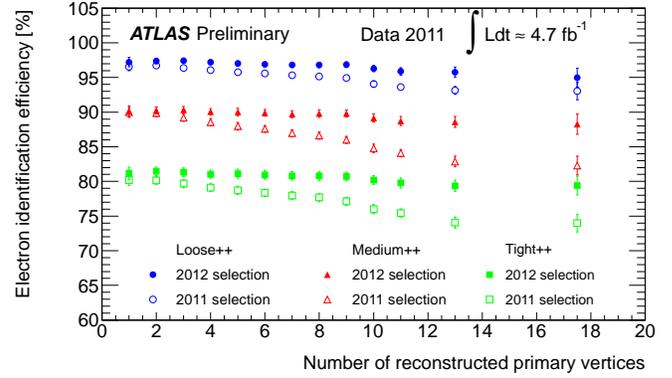}
\caption{Dependence of electron identification efficiency on pileup, measured using
$Z\rightarrow ee$, $J/\psi\rightarrow ee$ and $W\rightarrow e\nu$ events~\cite{Bocci:1403067}.}
\label{fig:electronEff}
\end{figure}

In ATLAS, electron identification~\cite{ATL-PHYS-PUB-2011-006} is performed with a cut-based method
combining information from the shower shape characteristics in the calorimeter with the information
from the ID where available.  Three reference sets of
cuts have been defined with increasing
background rejection power: \textit{loose}, \textit{medium} and \textit{tight} with an expected
jet rejection of approximately 500, 5000 and 50000, respectively, based on MC simulation. The values
of the cuts are optimized according to the beam conditions, leading to an improved set of cuts for
2012 which performs better in high pileup conditions while maintaining similar jet rejection power.
As can be seen in Fig. \ref{fig:electronEff}, an identification efficiency exceeding 95\%, 88\% and
79\% is achieved in 2012 for \textit{loose}, \textit{medium} and \textit{tight} respectively, for
up to 20 reconstructed primary vertices per event. 

\begin{figure}[!t]
\centering
\includegraphics[width=3.8in]{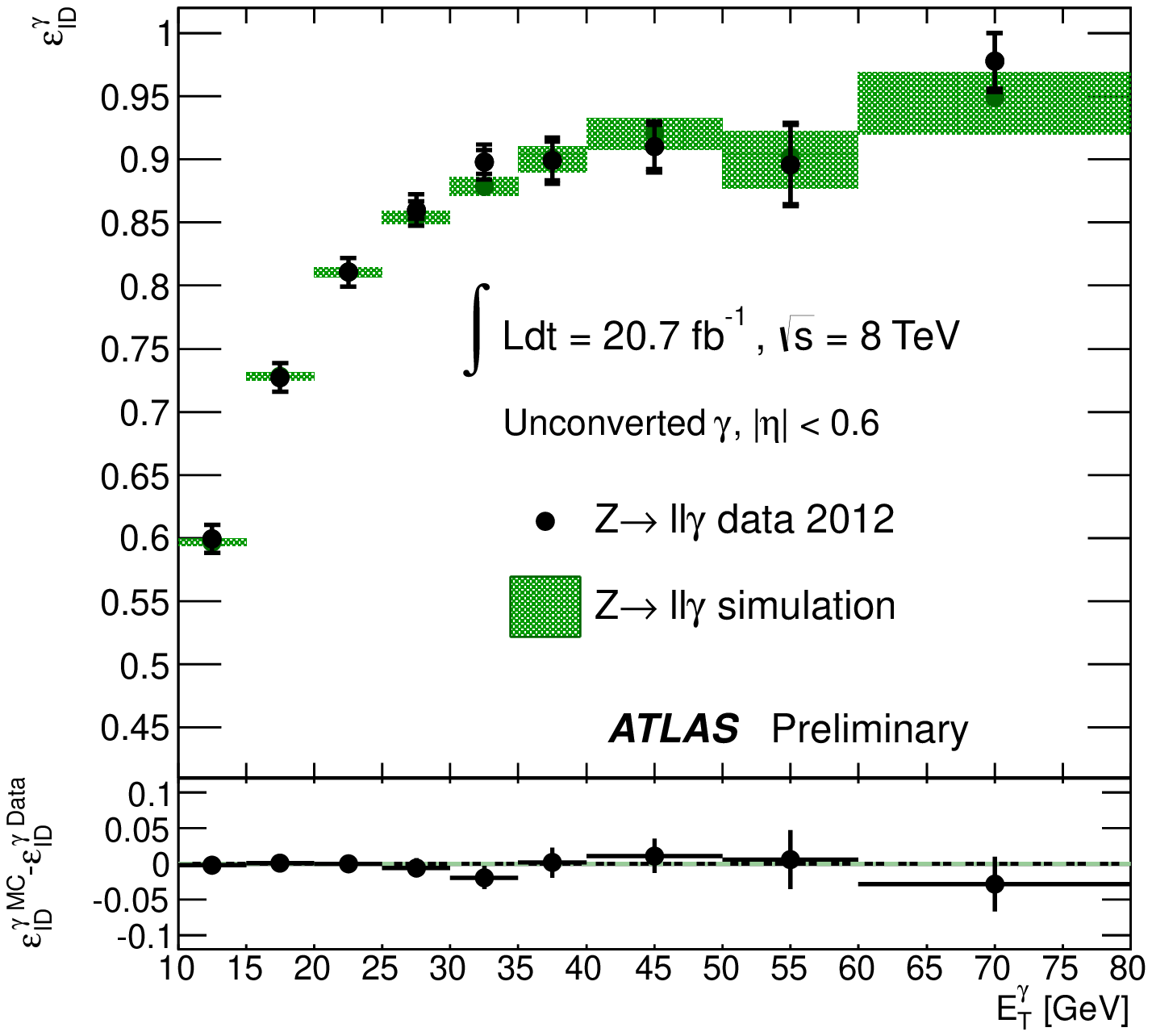}
\\
\includegraphics[width=3.8in]{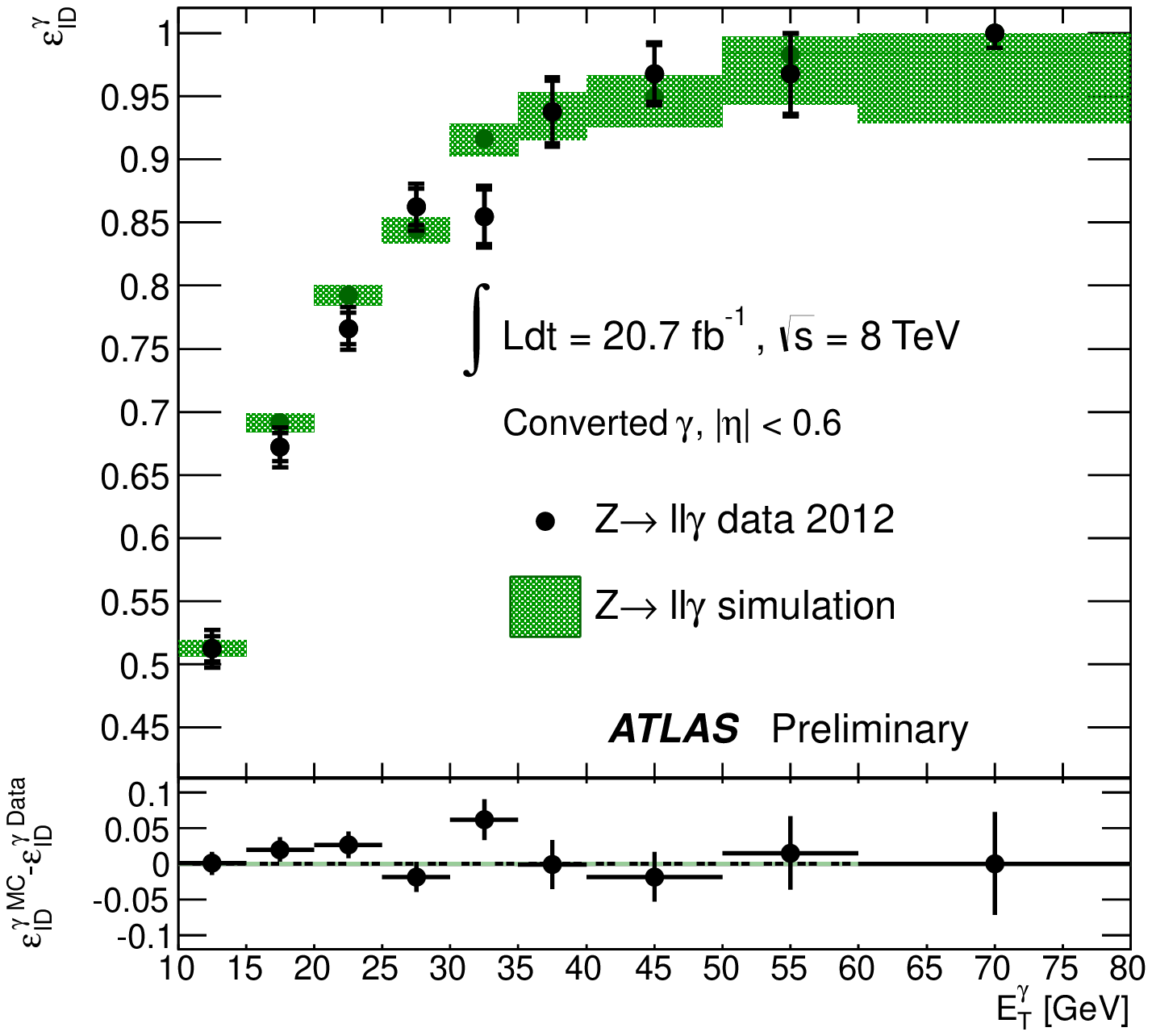}
\caption{Photon identification efficiency for unconverted (top) and converted
(bottom) photon candidates obtained in $Z\rightarrow \ell\ell\gamma$ events, for $|\eta|<0.6$.
Similar
performance is observed in other $\eta$ ranges~\cite{Bocci:1521320}.}
\label{fig:PhotonIdEff}
\end{figure}

Photon reconstruction~\cite{ATL-PHYS-PUB-2011-007} is similar to electrons albeit more
involved due to the fact that photons can be either unconverted or converted. The latter
are characterized by the presence of at least one track in the ID that matches the EM cluster in
the calorimeter, resulting in an ambiguity in the distinction between converted photons and
electrons. In addition, unconverted photons can also be reconstructed as electrons if their EM
clusters are erroneously associated with tracks that typically have low momentum. For this reason,
a procedure has been established to recover photon candidates from a collection of electron
candidates, based on combined information from the ID and the calorimeter (number and momentum of
matched tracks, number and position of hits in the ID, $E/p$ ratio). To provide a pure
photon sample and study the photon identification efficiency, the radiative $Z$ decay $Z\rightarrow
\ell\ell\gamma$ is exploited~\cite{Bocci:1521320}, where the lepton $\ell$, can be an electron or a
muon. In these
studies, no requirements on the photon shower shape are applied, to avoid any bias. As shown in
Fig. \ref{fig:PhotonIdEff}, the efficiency is very high  for both converted and unconverted
candidates, and agrees with the expected performance from MC simulation.

\begin{figure}[!t]
\centering
\includegraphics[width=3.5in]{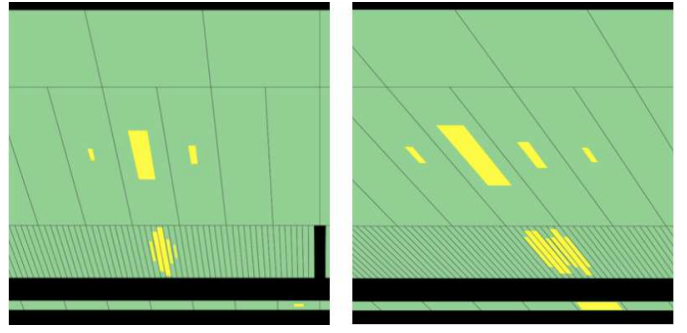}
\caption{Shower shapes for a photon candidate (left) and a candidate for a jet with a leading
$\pi^0$ (right), in data recorded in proton-proton collisions~\cite{displayPhotons}.}
\label{fig:photonVsPi0}
\end{figure}

The fine granularity of the calorimeter is a significant asset allowing the separation of photons
from jets. As can be seen in Fig. \ref{fig:photonVsPi0}, the shower shape for a photon is expected
to have a narrower profile compared to the shower shape for a jet. Especially in the presence of a 
$\pi^0$ meson, a distinctive energy deposition with two energy maxima is expected in the first layer
of the LAr calorimeter (strips). To efficiently reject background in analyses using photons, photon
identification is performed, similarly to electrons, using the characteristics of their shower
shape. Two reference sets of cuts, \textit{loose} and \textit{tight}, are defined. 
The former set of cuts has a very high efficiency with a modest jet rejection power, while the
latter has  a rejection power of approximately 5000 while keeping a relatively high efficiency,
approximately 85\% for photons with transverse energy $E_{\mathrm{T}}> 40$~GeV
\cite{ATLAS-CONF-2012-123}. 
 
\subsection{The Discovery of a Higgs-Like Boson in ATLAS}
A high functionality of all detector components is required for all ATLAS physics analyses.
Especially in the case of searches for new physics, the efficient running of the detector as a whole
is imperative for the rapid accumulation of data delivered by the LHC. As described in previous
sections, the LAr calorimeter operated close to optimally over the last three years, contributing to
the
very high ATLAS overall efficiency which culminated in the discovery~\cite{Aad:2012tfa} of a
Higgs-like boson in July 2012. By providing efficient identification and precise measurement of
electrons and photons, the LAr calorimeter was invaluable in the search
for the Higgs boson especially in its decay channels ${H\rightarrow2\mu2e}$, ${H\rightarrow4e}$
\cite{ATLAS-CONF-2013-013} and ${H\rightarrow\gamma\gamma}$~\cite{ATLAS-CONF-2013-012}.

\begin{figure}[!t]
\centering
\includegraphics[width=3.5in]{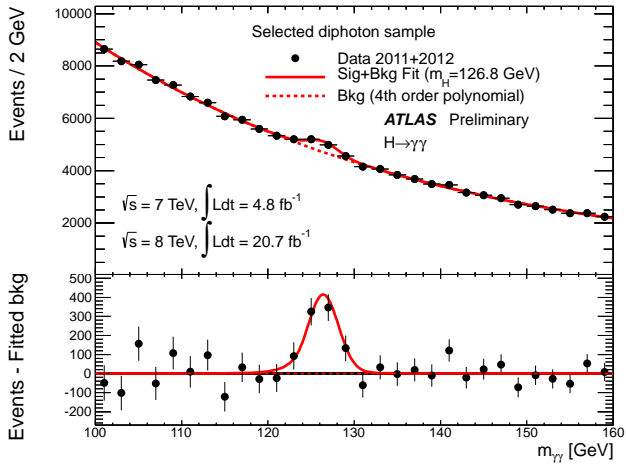}
\caption{Invariant mass distribution of diphoton candidates for the combined $\sqrt{s} = 7$~TeV and
$\sqrt{s} = 8$~TeV data samples. The result of a fit to the data of the sum of a signal component
fixed to $m_H = 126.8$~GeV and a background component described by a fourth-order Bernstein
polynomial is superimposed. The bottom inset displays the residuals of the data with respect to the
fitted background component~\cite{ATLAS-CONF-2013-012}. }
\label{fig:hgg}
\end{figure}

\begin{figure}[!t]
\centering
\includegraphics[width=3.5in]{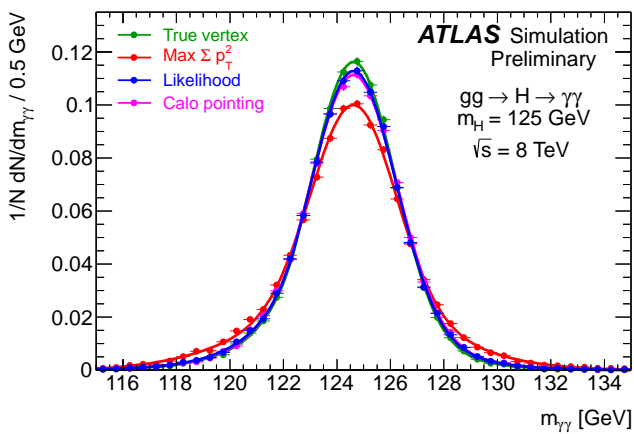}
\caption{Distribution of the expected diphoton mass for $H\rightarrow\gamma\gamma$ signal events as
a function of the algorithm used to determine the longitudinal vertex position of the
hard-scattering event~\cite{ATLAS-CONF-2012-091}.}
\label{fig:caloPointing}
\end{figure}

In the search for the Higgs decay to two photons, the LAr calorimeter
played an especially significant role. In brief, the $H\rightarrow\gamma\gamma$ analysis searches
the diphoton invariant mass spectrum for a small excess over a formidable background (Fig.
\ref{fig:hgg}). As described in
Section \ref{sec:id}, the calorimeter provides excellent photon identification and jet background
rejection, by exploiting its fine longitudinal segmentation, thereby improving the signal to
background ratio. Further, the diphoton invariant mass, defined as
\begin{displaymath}
 m_{\gamma\gamma} = \sqrt{2E_1 E_2 \left( 1- \cos\theta\right)}
\end{displaymath}
where $E_1$, $E_2$ are the two photon energies and $\theta$ is the angle between them, needs to be
reconstructed with a very high precision so a very small excess can be distinguished over a large
background. With an excellent energy resolution, the calorimeter ensures a very good measurement of
the photon energies. In addition, the unique capability to reconstruct the photon direction is a
significant advantage in improving the angle measurement in the high pileup environment of the LHC.
More specifically, the calorimeter pointing is used to choose the event primary vertex (PV) from
a set of several tens of PV candidates. As shown in Fig.
\ref{fig:caloPointing}, the use of the calorimeter information, labeled as ``Calo pointing'' is
significantly better than the ID-only approach, labeled as ``Max $\sum p_{\mathrm{T}}^2$'', which
selects the PV based on the momentum of the tracks associated with each candidate. Further, it is
very similar to the optimal achievable mass resolution labeled as ``True vertex''. Finally, the
``Likelihood'' method combines this information with the primary vertex information from the
tracking and provides similar mass resolution. With the PV position known with high precision
from the ID, the angle between the two photons is calculated with a very good resolution,
which translates to an excellent invariant mass resolution.

\section{Upgrade Plans}
The High Luminosity (HL) LHC intends to extend and improve the LHC program beyond 2022, by
operating with an instantaneous luminosity up to 5 times the original design 
luminosity of $\mathrm{10^{34}~cm^{-2}s^{-1}}$. Already by 2012, the LHC has been operating very
close to the design luminosity and after the ongoing shutdown (LS1), in 2015 it is intended to
exceed it. After an additional shutdown period in 2018 (LS2), the accelerator chain will be
upgraded so the LHC can reach a luminosity of $3\times10^{34}~\mathrm{cm^{-2} s^{-1}}$. A road map
has been set to upgrade the calorimeter in two phases, as the HL--LHC era approaches, to maintain a
high level of performance in these harsh conditions. 

In Phase--1~\cite{CERN-LHCC-2011-012}, during LS2, the calorimeter L1 trigger will be
upgraded to cope with the high rates expected. Given a limited bandwidth allowance, it is otherwise
required to either raise the trigger thresholds or randomly discard interesting events, neither of
which is desired. The proposed solution is to upgrade the existing trigger electronics with a new
front-end LAr Trigger Digitizer Board and additional back-end electronics. In this architecture,
the coarse granularity trigger towers will be complemented with finer granularity ``super-cells''.
In addition, it will be possible to perform complex shower shape calculations at the hardware
level. These new features allow better rejection of undesired background events online using
characteristics of the shower shape, in a similar way as is done in physics analyses offline,
therefore saving the bandwidth for more interesting events.

For the Phase--2 upgrade~\cite{ATLAS:1502664}, the
front-end electronics will move to a fully digital architecture~\cite{Andeen:1476916} and will take
advantage of the technology developed for the Phase--1 trigger upgrade. In the proposed
architecture, the signal will be digitized continuously at the 40~MHz LHC clock frequency and
transmitted to an off-detector back-end system able to handle readout at 150~TBps. At the back-end,
input will be provided to a new fully-digital L1 trigger system. It is also proposed to use the
trigger upgrades from Phase--1 to create an additional, lower level for triggering, dubbed Level--0.

Finally, in the HL--LHC environment, degradation of the FCal performance is expected
\cite{Rutherfoord:1473138}. The increased particle fluxes will have the following effects to FCal
operation: First, at high ionization rates, space-charge effects become significant, when the
positive argon ions build up in the LAr and distort the electric field. Secondly, the high
currents expected will cause a large voltage drop over the HV protection resistors in place to
protect the system from arcing. Both these effects degrade the signal measurement. Finally, the
large rates and currents may generate heat to the level that can boil the LAr
which is of course undesirable. 

A first option is to replace the current FCal modules with ones able to withstand the higher fluxes
anticipated. The new modules would feature smaller LAr gaps to solve the
space-charge issue while the value of the protective resistors would be reduced to avoid a
significant voltage drop with the currents expected. Finally, they would employ liquid nitrogen
cooling loops to remove the excess heat. This option would require the opening of the cryostat which
is projected to be a challenging and risky operation. This solution would therefore be especially
considered should there be a particular need to open the cold vessel. For example, unlike the other
LAr subsystems, the HEC employs cold GaAs pre-amplifiers that are located inside the endcap
cryostats in LAr. Studies are underway to determine whether the HEC cold electronics will need to be
replaced for HL--LHC, in which case the FCal will need to be removed from the cryostat to allow
access to the HEC. Due to activation of the FCal, it will not be
possible to re-install it after the HEC intervention, without exposing the personnel to
unacceptable radiation levels. In this scenario, therefore, complete replacement of the modules is
favored. 

A second option would consist of installing an additional warm calorimeter (named
\textit{miniFCal}) in available space in front of the existing FCal system, closer to the
interaction point. This option is attractive since it avoids the risk of opening the cryostat
cold volume and can be installed more readily, thereby saving time. Various technologies have been
proposed for the miniFCal. One of them, very radiation hard, involving diamond wafers as active
material, has been studied to assess its linearity, uniformity and resolution. However, this option
is disfavored mainly due to its prohibiting cost. Other considered
technologies include xenon gas as active medium  as well as familiar LAr technology, each with
their own implementation challenges. 

Finally there is the option of not taking any action and living with the degraded performance and
the possibility of the LAr boiling at very high rates, which has yet to be explicitly ruled out.
Additional studies are required to better quantify the level of FCal degradation to be anticipated
in HL--LHC, before a decision is made taking into account detector safety, schedule and cost.

\section{Conclusion}
The LAr calorimeter system has performed exceptionally well over the first few years of operation,
with near optimal efficiency as well as impressive performance, contributing to the overall ATLAS
performance that culminated in the discovery in 2012 of the long
sought after Higgs boson. Keeping this level of performance requires a tremendous effort by a
significant number of people who intend to keep or exceed this level in the coming challenging
years. Studies are continuing in order to improve the performance as the LHC delivers data at
unprecedented energies and rates, and ensure that any physics that exists at the TeV-scale will be
discovered.


%



\bibliographystyle{IEEEtran}
\bibliography{IEEEabrv,paper}
\end{document}